\documentclass[preprint2]{aastex}
\usepackage{graphics}
\usepackage{psfig}
\usepackage{longtable}
\usepackage{psfig}
\def\gsim{\lower 2pt \hbox{$\, \buildrel {\scriptstyle >}\over
{\scriptstyle \sim}\,$}}
\def\lsim{\lower 2pt \hbox{$\, \buildrel {\scriptstyle <}\over
{\scriptstyle \sim}\,$}}

\def\rosat{{\sl ROSAT}}

\def\chandra{{\sl Chandra}}

\newcommand{\as}{$^{\prime\prime}~$}

\def\xs{{Abell 2125}}

\shortauthors{}
\shorttitle{}

\begin{document}
\title{X-raying Abell 2125 ---  a Large-scale Hierarchical Complex of Galaxies and Hot Gas}
\author{Q. Daniel Wang\altaffilmark{1}, Frazer Owen\altaffilmark{2}, 
and Michael Ledlow\altaffilmark{3}}
\altaffiltext{1}{Department of Astronomy, University of Massachusetts, 
Amherst, MA 01003}
\altaffiltext{2}{National Radio Astronomy Observatory, P.O. Box O, Socorro, NM 87801, USA}
\altaffiltext{3}{Gemini Observatory, Southern Operations Center, AURA, Casilla 603, La Serena, Chile}
\shortauthors{Wang et al.}
\shorttitle{{\sl Chandra} Observation of Abell 2125}


\begin{abstract}
We present an 82 ksec {\sl Chandra} ACIS-I observation of a large-scale 
hierarchical complex, which consists of various clusters/groups of galaxies 
and low-surface brightness X-ray emission at $z = 0.247$. This high-resolution 
{\sl Chandra} observation allows us for the first time to separate 
unambiguously the X-ray contributions from discrete sources and large-scale
diffuse hot gas. We detect 99 X-ray sources in a $17^\prime \times 17^\prime$
field. Ten of these sources are identified as members of the complex and 
are mostly radio-bright. Whereas unresolved X-ray sources tend to 
be associated with galaxies in intermediate density environments, extended 
X-ray emission peak at bright radio galaxies in the 
central cluster. In particular, a distinct X-ray trail appears on one 
side of the fast-moving galaxy C153, clearly due to ram-pressure stripping. 
The diffuse X-ray emission from the 
central cluster can be characterized by a thermal plasma with a 
characteristic temperature of $3.2_{-0.4}^{+0.5}$ keV and a heavy 
element abundance of $0.24_{-0.12}^{+0.15}$ solar (90\% confidence 
uncertainties). In comparison, a patch of low-surface 
brightness X-ray emission apparently originates in
relatively low density intergalactic gas with a characteristic temperature 
of $0.98_{-0.27}^{+0.22}$ keV and an abundance of $\lesssim 0.09$ solar. 
The {\sl Chandra} observation, together with extensive multi-wavelength
data, indicates that the complex represents a projection of several 
galaxy sub-structures, which may be undergoing major mergers.
We discuss the dynamic states of the complex and its sub-structures as
well as properties of X-ray-emitting galaxies and the relationship to their 
environments.
\end{abstract}


\section{Introduction}

The structure of the universe is believed to have formed by clustering 
hierarchically from small to large scales. The outcome of this hierarchical 
formation process depends largely on the interplay between galaxies and 
their environments. But how and where such galaxy-environment interactions 
primarily occur 
remain greatly uncertain (e.g., David et al. 1996; Wang \& Ulmer 1997; Ponman et al.
1999; Fujita 2001; Balogh et al. 2002; Bekki et al. 2002).

        We have identified a large-scale hierarchical complex (Fig.\ 1)
that is well-suited for investigating the structure formation process and
the environmental impact on galaxy properties. Revealed in a survey of 
10 Butcher \& Oemler clusters observed with the \rosat\ PSPC (Wang 
\& Ulmer 1997), this complex 
contains various X-ray-emitting features, which are associated with
concentrations of optical and radio galaxies (Fig.\ 1; 
Wang, Connolly, \& Brunner 1997a; Owen et al. 1999; Dwarakanath \& Owen 1999). 
The overall optical galaxy concentration of the region
was originally classified as a cluster Abell 2125 (richness 4). 
The \rosat\ image and follow-up optical observations, 
however,  have shown that the complex contains three well-defined 
X-ray bright clusters (Wang et al. 1997a). In addition, substantial amounts of 
unresolved low-surface brightness X-ray emission (LSBXE) are also present.
The overall angular size of the entire X-ray-emitting complex seems to 
extend more than $\sim 30^\prime$. But the three main concentrations of 
galaxies (LSBXE, the central A2125 cluster, and Cluster B) are 
identified within a smaller projected region
of dimension $\sim 12^\prime$ ($1^\prime = 0.23$ Mpc; the cosmological 
parameters, $H_0 = 71 {\rm~km~s^{-1}~Mpc^{-1}}$, $\Omega_{total} = 1$, and
 $\Omega_\Lambda = 0.73$ are adopted throughout the paper).  
The complex thus represents an X-ray-bright hierarchical filamentary 
superstructure, as predicted by numerical simulations of the structure 
formation (e.g.,  Cen \&  Ostriker 1996). 


\begin{figure}[htb!]
\caption{\protect\footnotesize
\rosat\  PSPC X-ray image of the Abell 2125 complex and its vicinity in the 
0.5-2~keV band (Wang et al. 1997a). Point-like X-ray sources detected in the
image have been excised. Each contour is 50\% (2$\sigma$) above 
its lower level. Spectroscopically confirmed, radio detected members of 
the complex (Owen et al. 2004a) are marked by {\sl pluses}. The box 
outlines the field covered by our \chandra\ ACIS-I observation (Fig. 2).
The large cross and circle show the center position and 
radius used in the Abell catalog.}
\label{fig1}
\end{figure}

\begin{figure*} 
\unitlength1.0cm
\centerline{\vbox{
\vspace{-2cm}
\vspace{-2.2cm}
}}
\caption{\protect\footnotesize
\chandra\ ACIS-I images of the Abell 2125 field in
the 0.5-2 keV (upper panel) and 2-8 keV (lower panel) bands.
The images are smoothed with a 
Gaussian of FWHM equal to 3\as. The small circles represent the
regions removed for discrete sources, which are labeled as in Table 1.
The two large circles represent the cluster and the LSBXE,
while the large ellipse outlines the diffuse emission to the
southwest. The two squares outline the off-source background regions, 
from which spectra are extracted. 
}
\label{fig2}
\end{figure*}

The nomenclature adopted here  (e.g., in Fig. 1) follows that used 
in Wang et al. 
(1997), based on X-ray identifications.   We call the entire elongated 
diffuse X-ray enhancement as seen in Fig. 1 as the Abell 2125 complex.
Historically, however, this congregation of optical galaxies was loosely called
a cluster. The centroid listed in the Abell catalog (Abell 1958) is 
R.A., Dec. (J2000) = $15^h40^m55^s$, 66$^\circ 19^\prime15^{\prime\prime}$,
as marked in Fig. 1.  The Abell radius, 9\farcm2 (2 Mpc),  
encloses the bulk of the galaxy concentrations in the field of Fig. 1, 
including Cluster B and much of the LSBXE. The work by Butcher et al.
(1983) and Butcher \& Oemler (1984), however, assumed a cluster 
center that is close to 
what is inferred from the X-ray for 
the central Abell 2125 cluster (Fig. 1).
Furthermore, their imaging field of view (FoV $= $ 55 arcmin$^2$)
is comparable to the size of the X-ray cluster, but is much smaller than 
the area enclosed in the Abell radius.
The central cluster itself, which is only part of the complex, 
is not particularly rich. This can, 
at least partly, explain its relatively large blue galaxy 
fraction ($f_b \sim 20\%$; Butcher \& Oemler 1984).
But, the Abell 2125 complex does seem to contain a distinct population of 
active galaxies (Owen et al. 2004a). There is an over-abundance
of radio galaxies in Abell 2125, $\sim 9\%$ compared to 
a typical 2\% for rich clusters at similar redshifts (Morrison \& Owen
2003).

We have obtained a deep \chandra\ observation that covers part of the Abell 2125 
complex (Fig. 1) 
to characterize its detailed X-ray properties and to study the 
interplay between galaxies with their environments.
The high spatial resolution of \chandra\
enables us to examine diffuse X-ray structures down 
to a scale of $\sim 3.8$~kpc ($\sim 1^{\prime\prime}$). In this paper, we 
concentrate on presenting the observation and the results on the
detection of discrete sources and on the characterization of large-scale
diffuse X-ray emission. Detailed analysis of the X-ray data in
conjunction with 
observations in other wavelength bands will be discussed elsewhere 
(e.g., Owen et al. 2004a,b), although a few galaxy counterparts 
 will be mentioned in the present work. X-ray sources are labeled with
a two-digit number (with a prefix ``X-'' in the text), whereas optical 
IDs are in five-digits (e.g., 00047).

\section{Observation and data Analysis}

The \chandra\ observation of Abell 2125  (Obs. ID. 2207) was taken on Aug. 24,
2001 for an exposure of 81.6 ks. The instrument ACIS-I was at the 
focal plane of the telescope and in the ``VERY FAINT'' mode.
Our data calibration follows the same procedure detailed 
by Wang et al. (2004). Briefly,
we reprocessed the level 1 (raw) event data to generate a new level 2 event
file, using the \chandra\ Interactive Analysis of Observations software package
(CIAO; version 2.3). 
The re-processing includes an improved absolute astrometry
of the observation (nominally better than $\sim 0\farcs3$), a correction
for the charge transfer inefficiency of the CCDs, and a better software 
rejection of particle-induced events (i.e., setting the flag 
check$\_$vf$\_$pha=yes when running the CIAO program acis$\_$process$\_$events). We further removed time 
intervals with significant background flares, or peaks with count rates
$\gtrsim 3\sigma$ and/or a factor of 
$\gtrsim 1.2$ off the mean background level of the observation, 
using Maxim Markevitch's light-curve cleaning routine ``lc$\_$clean''. 
This cleaning, together with a correction for the dead time of the 
observation, resulted in a net 80.5 ks exposure (livetime) 
for subsequent analysis. 

We constructed broad band X-ray intensity images by flat-fielding 
count images with corresponding effective exposure maps, weighted
with a power law of photon index 1.7. This flat-fielding corrected for 
telescope vignetting and bad pixels as well as the 
quantum efficiency variation of the instrument, 
including an observing date-dependent sensitivity degradation, which is 
particularly important at low energies $\lesssim 1.5$ keV. 

We searched for X-ray sources in the three broad bands, (S) 0.5-2,  (H)
2-8, and (B) 0.5-8 keV. In each band, we successively applied the source 
detection and analysis algorithms: wavelet, sliding-box, and maximum 
likelihood centroid fitting; the procedure for these applications as well
as the related background map construction was detailed by 
Wang et al. (2003). The significance of a source detection
is characterized by the false detection probability $P$ due to a random 
background fluctuation:
\begin{equation} 
P =1- \sum_{n=0}^{n_c-1} {n_b^n \over n!} e^{-n_b},
\end{equation} 
where $n_c$  and  $n_b$ are the total numbers of counts, 
detected and expected from the local diffuse 
background. The above definition of $P$ is 
slightly different from that used in Wang et al. (2003), where the upper limit
of the summation in equation (1) was set to be $n_c$.
When  $P < 10^{-6}$ (our chosen threshold), 
a positive source detection was claimed. 
Source count rates were estimated within the 90\% 
energy-encircled radius (EER) of the point spread function (PSF), which is off-axis
dependent (Fig. 2; Jerius et al. 2000). The calculation of the EER  assumed
a mean X-ray spectrum as a power law with a photon index equal to 1.7, typical
for an AGN. But individual sources could have very different spectra.
To minimize the dependence on the assumed spectrum, the correction 
was made individually in each of the four narrower 
bands (0.5-1, 1-2, 2-4, and 4-8 keV), before
the source count rates were combined into the
above broad bands. 
The spectral variation also introduce uncertainties in our estimate of the 
detection threshold. This explains why a couple of 
sources have count rates below the thresholds in Fig. 3.

\begin{figure} 
\centerline{ {\hfil\hfil
\psfig{figure=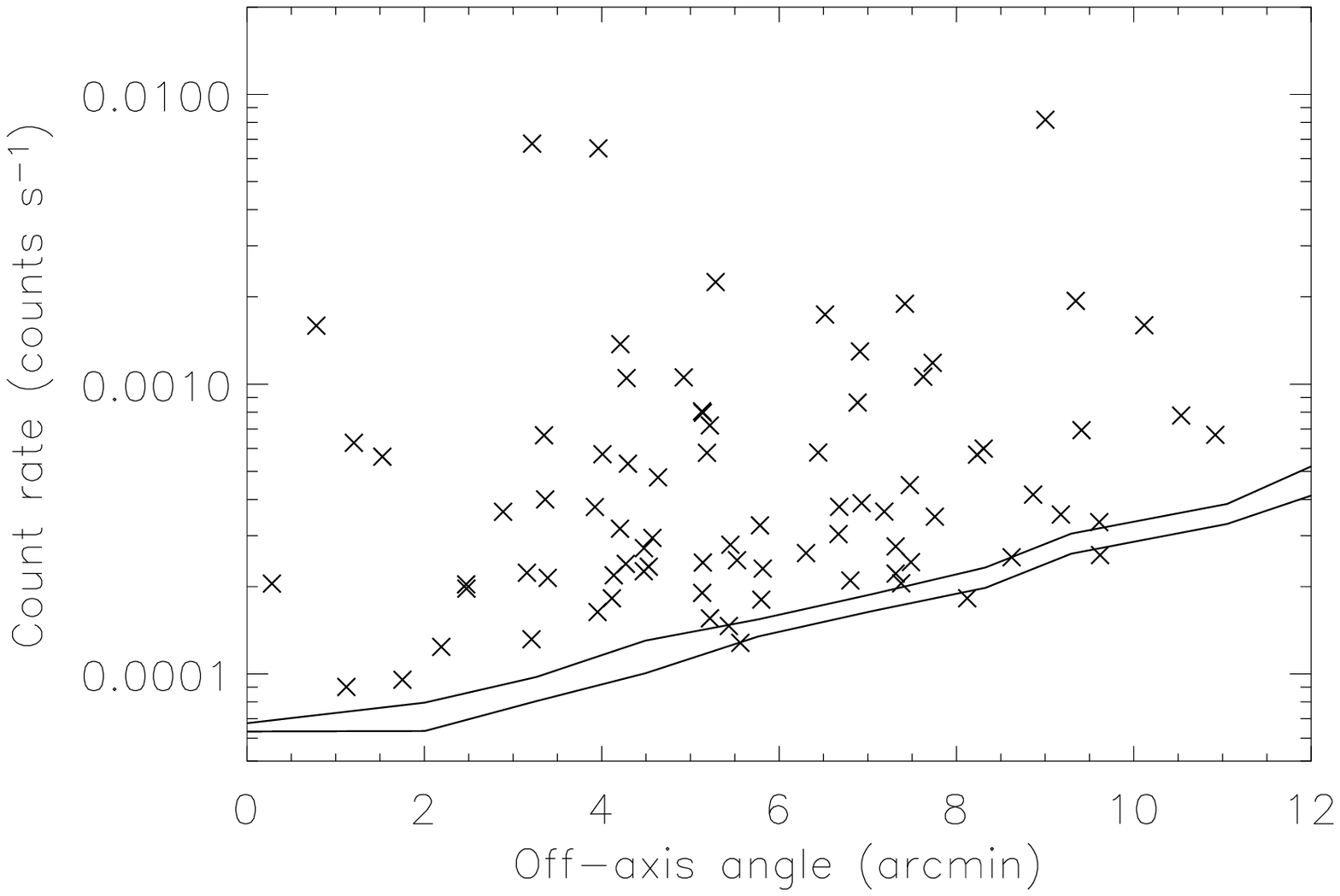,height=2.7in,angle=0, clip=}
\hfil\hfil}}
\caption{\protect\footnotesize
Count rates of sources detected in the 0.2-5 keV band
vs. the off-axis angle of the ACIS-I observation.
The upper and lower curves illustrate
the approximate 0.2-5 keV band detection thresholds, 
the azimuthally averaged and the lowest at each off-axis angle, respectively. 
}
\label{fig3}
\end{figure}

To determine how many of our detected sources may be associated with the A2125 complex, we need to 
estimate the number of possible interlopers  (e.g., background AGNs).
We use the Log($N$)--log($S$) relation
presented by Moretti et al. (2003 and references therein) for sources
detected separately in the 0.5-2 keV and  2-10 keV bands,
assuming a power law of photon index of 1.4 and a foreground
absorption column density $N_H = 1.6 \times 10^{20} {\rm~cm^{-2}}$.
To convert the flux $S$ back into the ACIS-I count rate, 
we assume the same
power law and $N_H = 3 \times 10^{20}  {\rm~cm^{-2}}$, appropriate for
our field. The corresponding conversion is $0.45 $ and 
 $3.0 \times 10^{-11}$ ${\rm~(erg~cm^{-2}~s^{-1}})/({\rm counts~s^{-1}})$
in the 0.5-2 keV and 2-10 keV bands, respectively. 

The sensitivity of our source detection depends on the size of the PSF as 
well as the local background level and effective exposure, 
which all vary with position, especially with the off-axis angle (Fig. 3). This variation can seriously affect the number-flux distribution
of the detected sources, except for those with count rates $\gtrsim 5 \times 10^{-4}
{\rm~counts~s^{-1}}$ (Fig. 3). Furthermore, because of the limited
counting statistics, many sources have very uncertain count rates. 
As a result of this large uncertainty, together with a typical
steep number-flux distribution of background sources, 
more faint sources are expected to be scattered to fluxes above the detection 
limit than the other way around. This is similar to the Eddington 
bias in optical photometry (Hogg \& Turner 1998) and is accounted for
in our calculation of the expected interlopers in the field, according to
the procedure detailed in Wang (2004). 

We removed a region of twice the 90\% 0.5-8 keV band EER around each source to 
study diffuse X-ray emission (Fig. 2). For visual presentation of 
smoothed diffuse X-ray intensity maps, we replaced the source-removed region
with values estimated from data in surrounding bins.  

In our diffuse X-ray spectral analysis, we take  
special care of the background subtraction. We estimated 
the background contribution in off-source regions (Fig. 2). To check
the consistency of the background spectral 
properties in different regions, we compared the on-
and off-source spectra extracted from the so-called blank-sky 
data\footnote{available at 
http://cxc.harvard.edu/contrib/maxim/acisbg/data/README}. With
an effective livetime of 550 ks, the data were re-projected to mimic our 
observation,
accounting approximately for both the cosmic X-ray background and the 
contribution from charged particle-induced events. 
We found that a simple re-normalization of the off-source spectrum 
yielded statistically consistent on- and off-source 
blank-sky spectra. The re-normalization corresponded to 
a reduction of the livetime of the off-cluster spectrum by a factor of 1.16
 ($\chi^2/d.o.f. = 360.7/328$).
The renormalized off-source background of our observation was then used for
the background subtraction in spectral analysis within XSPEC (Arnaud 1996).
For the LSBXE feature, the on- and off-source blank-sky spectra, are 
statistically consistent with each other ($\chi^2/d.o.f. = 482.5/450$). The
best-fit normalization factor is 1.02, resulting only marginal improvement of
the fit ($\chi^2/d.o.f. = 472.5/450$). We
therefore, used the off-source spectrum from our observation 
without the normalization for the background subtraction of the LSBXE 
spectral analysis. Similarly, no background intensity normalization is
needed for the southwest diffuse feature (Fig. 2; the southwest hereafter;
\S 3.2)

\section{Results}

\subsection{Discrete X-ray Sources}

Fig. 2 presents an overview of the ACIS-I image and the detected X-ray 
sources, which are listed in Table 1. Various parameters
are defined in the note to the table. 
The conversion from a count rate to an
absorption-corrected energy flux depends on the source spectrum (Fig. 4). 
The hardness ratios,
compared with the models (Fig. 5) may be used to characterize the X-ray 
spectral properties of relatively bright sources. A 
typical value of the conversion is $\sim 8 \times 10^{-12}$ 
${\rm~(erg~cm^{-2}~s^{-1}})/({\rm counts~s^{-1}})$ in the 0.5-8 keV band,
appropriate  
for a power law spectrum of photon index 2 and an absorbing-gas 
column density $N_H \sim 1 \times 10^{21} 
{\rm~cm^{-2}}$. This conversion should be 
a good approximation (within a factor of 2) for 
 $\lesssim 3 \times 10^{21} {\rm~cm^{-2}}$ (Fig. 4). The corresponding 
conversion to a source-frame luminosity in the same band is 
$\sim 1.4 \times  10^{45}$ ${\rm~(erg~s^{-1}})/({\rm counts~s^{-1}})$ at the
distance of Abell 2125. One can then estimate from the observed count 
rate the star formation rate
as $ \sim 6.8 \times 10^{-41} L_{\rm 2 - 8 keV} {\rm~M_\odot~yr^{-1}}$ for starburst
dominant galaxies (Ranalli, Comastri, \& Setti 2003; Owen et al. 2004a).

\begin{figure} 
\centerline{ {\hfil\hfil
\psfig{figure=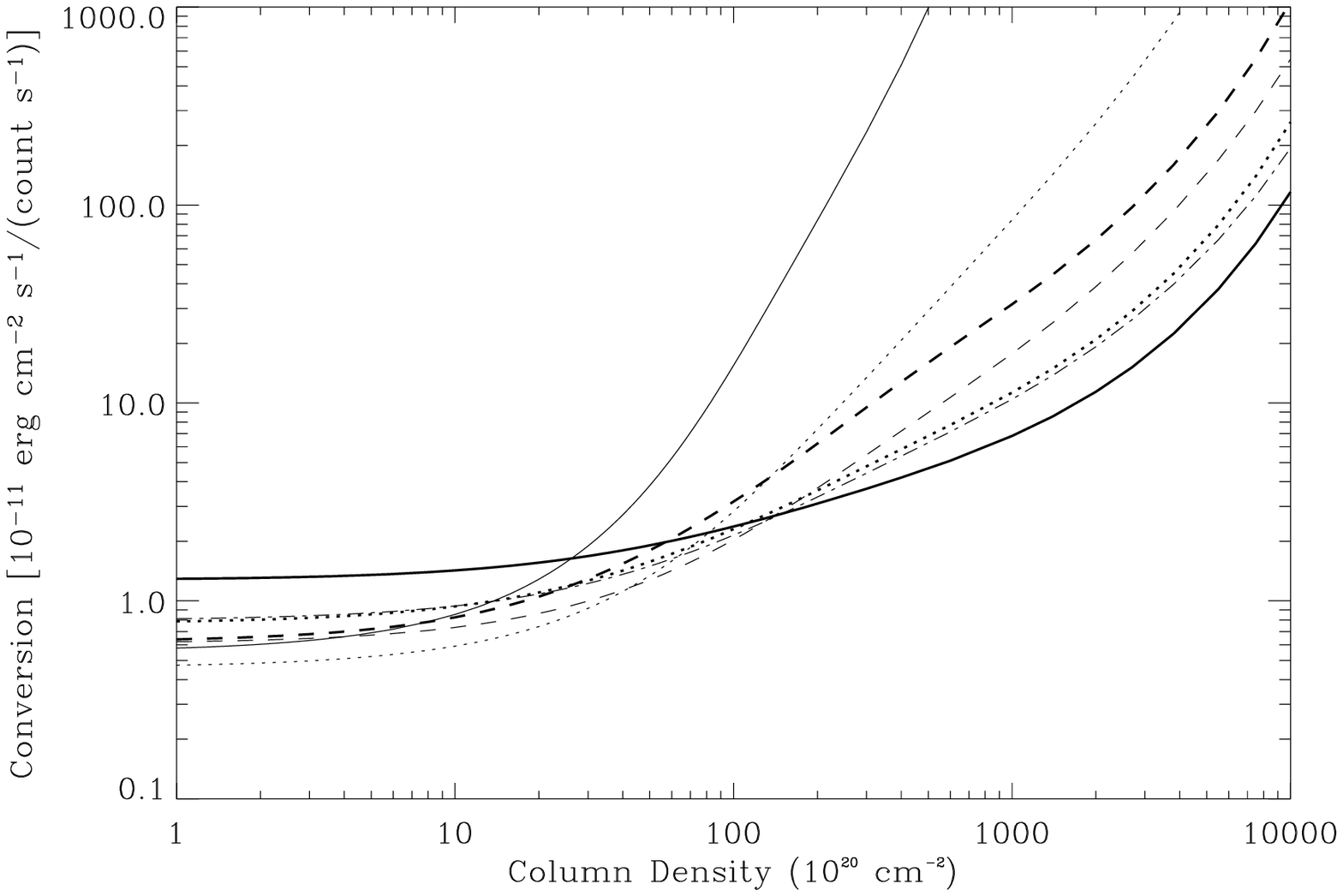,height=2.7in,angle=0, clip=}
\hfil\hfil}}
\caption{\protect\footnotesize
Conversion of an ACIS-I count rate to an energy flux  
in the 0.5-8 keV band for representative models: the
thick curves are for the power-law model
with a photon index equal to 1 (solid), 2 (dotted), and 3 (dashed), 
whereas the thin curves are for the thermal plasma with a temperature
equal to 0.3 (solid), 1 (dotted), 2 (dashed), and 4 (dot-dashed) keV, 
respectively. The heavy element abundance is assumed to be solar for
both the X-ray-emitting plasma and the absorbing gas.}
\label{fig4}
\end{figure}

\begin{figure*} 
\centerline{ {\hfil\hfil
\psfig{figure=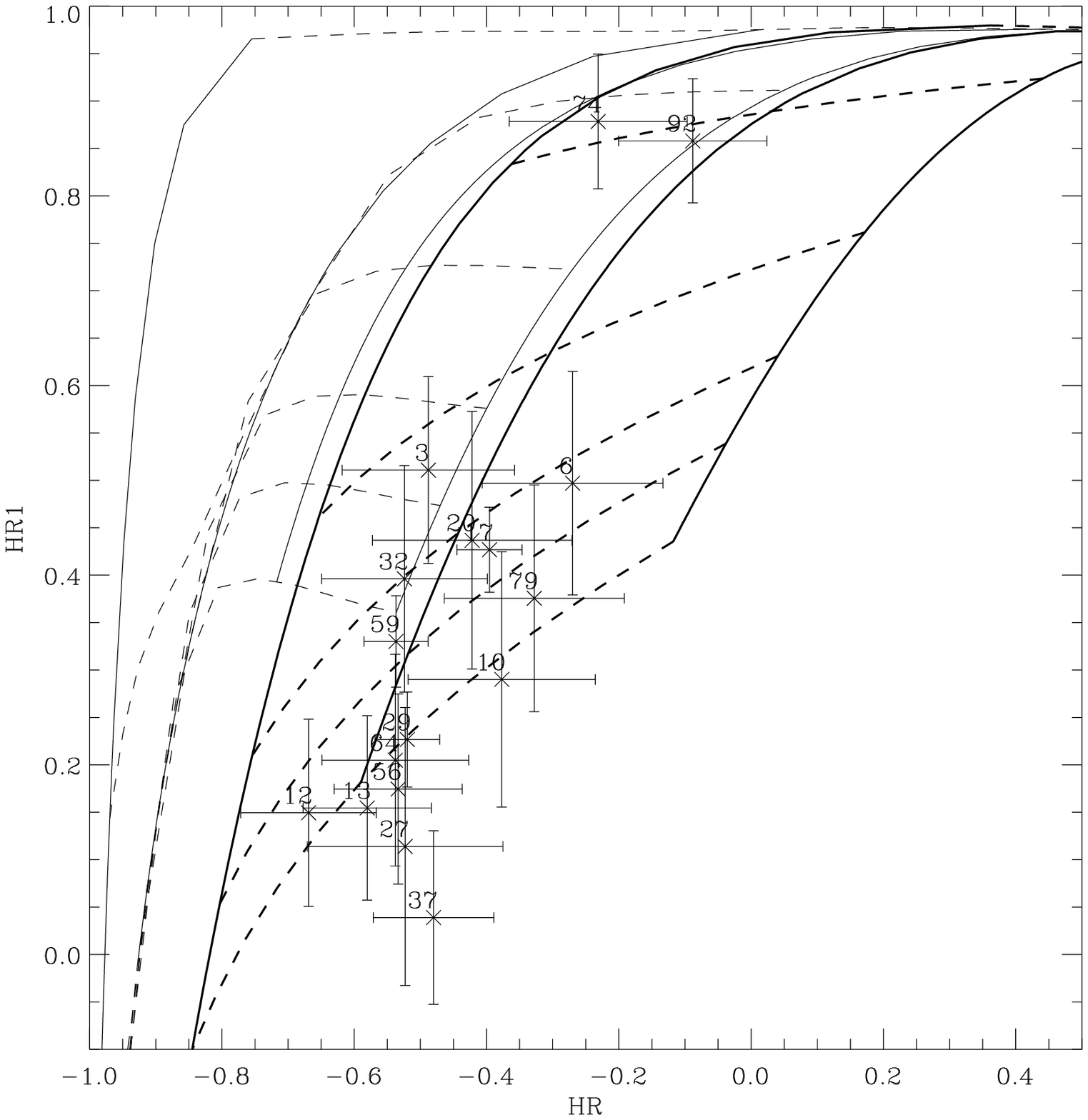,height=3.2in,angle=90, clip=}
\psfig{figure=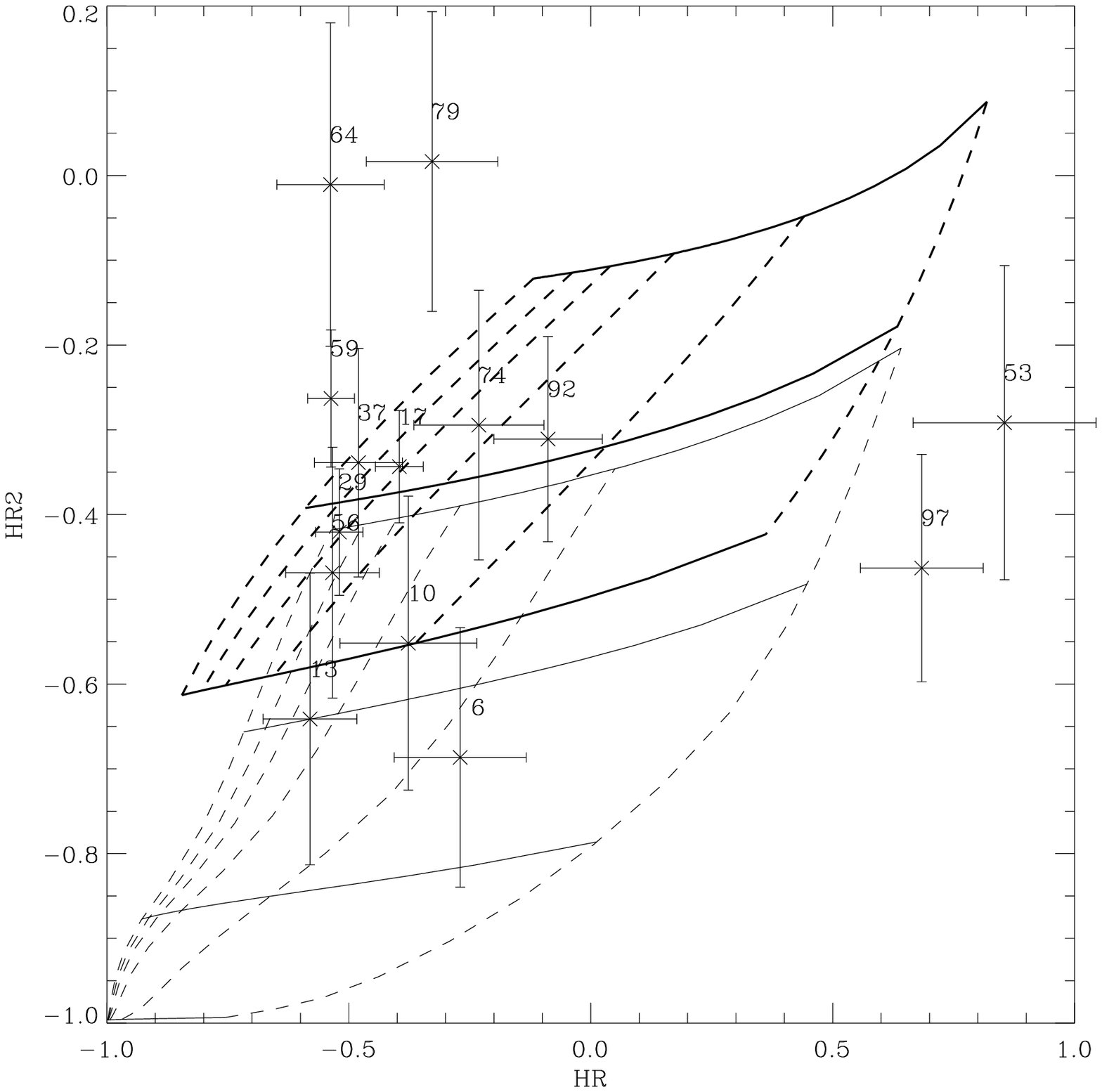,height=3.2in,angle=90, clip=}
\hfil\hfil}}
\caption{\protect\footnotesize
Color-color diagrams of X-ray sources with their generic 
numbers (Table 2) labeled. The hardness ratios (HR1,
HR2, and HR) are defined in the notes 
to Table 1, and the error bars represent 1$\sigma$ uncertainties. 
Also included in the plot are
hardness-ratio models: the solid thick curves are for the power-law model
with photon index equal to 3, 2, and 1, whereas the 
solid thin curves are for the thermal plasma with a temperature
equal to 0.3, 1, 2, and 4 keV, from left to right  in the left 
panel and from bottom to top  in the right panel, respectively. The absorbing
gas column densities are 1, 10, 20, 40, 100, and 300
$\times 10^{20} {\rm~cm^{-2}}$ (dashed curves from bottom to top in the left 
panel and from left to right in the right panel)}
\label{fig5}
\end{figure*}

\begin{figure} 
\centerline{ {\hfil\hfil
\psfig{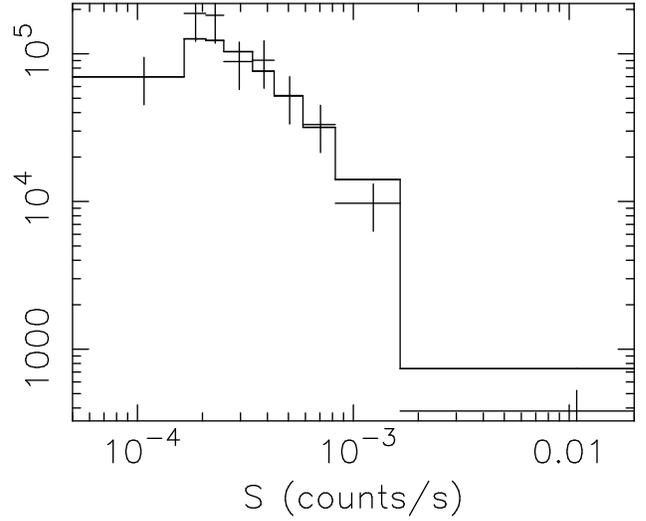}
\hfil\hfil}}
\caption{\protect\footnotesize
The source count rate distribution of detected X-ray sources,
excluding known complex members, compared with 
the expected background contribution (histogram) in the 0.2-2 keV band. }
\label{fig6}
\end{figure}

\begin{figure}
\unitlength1.0cm
\centerline{
}
\caption{\protect\footnotesize
ACIS-I intensity images of the Abell 2125 complex:
(a) a tri-color montage of X-ray intensities in the 0.5-2 keV (red), 
2-4 keV (green), and 4-8 keV (blue) bands; (b) the 0.5-2 keV band only with 
green {\sl squares} mark the positions of complex member 
galaxies (radio- and X-ray-bright ones are further marked in blue and with 
{\sl crosses}); (c) the 
core of the Abell 2125 cluster (2\farcm36$\times$2\farcm36 FoV) 
in the Kitt Peak 4-m mosaic V-band (red)
as well as in the 0.5 - 2 keV (green) and 2 - 8 keV  (blue) bands;
(d) a close-up of C153 (00047) in an {\sl HST} WFPC-2 V-band (red; Owen 
et al. 2004b) and its diffuse X-ray trail (the same color-code 
as in the panel c; 0\farcm426 $\times$ 0\farcm491 FoV).
}
\label{fig7}
\end{figure}

Most of the sources are detected 
in multiple bands, including 81 and 48 
detections in the 0.5 - 2 keV and 2 - 8 keV bands,
respectively.  In comparison, the
expected numbers of interlopers are 75 and 45 in the same two bands. 
Ten of the sources (Table 1) are in positional coincidence
with member galaxies of the A2125 complex
(Owen et al. 2004a). While one of the member sources (X-65) has a
count rate of $1.8  \times 10^{-3} {\rm~counts~s^{-1}}$, all others
have count rates $\sim 2.4-7.4 \times 10^{-4} 
{\rm~counts~s^{-1}}$, in which there is an excess of the observed
number-flux (count rate) distribution 
of sources above the predicted interloper contribution. With these
member sources excluded, which are   
all detected in the 0.5-2 keV band, the observed and predicted distributions 
are consistent with each other (Fig. 6). 
This direct comparison (without a fit) gives $\chi^2 = 9.9$ for 9 bins
compared with $\chi^2 = 20.5$ for 10 bins without the exclusion of the member 
sources. Therefore, the total excess above the expected 
number of interlopers is at $\sim 2$ confidence level, but few 
(if any) of the remaining unidentified sources are likely to be members 
of the complex.

\subsection{Diffuse X-ray Emission}

We present in Fig. 7 adaptively smoothed ACIS-I intensity 
images. The presence of the large-scale diffuse X-ray
emission is apparent. The overall morphology of the diffuse emission is
very similar to that seen in Fig. 1, although part of the PSPC counts 
is clearly due to point-like sources detected now in the ACIS-I image. 
The diffuse emission is substantially
softer than  typical discrete sources. There are substructures in the diffuse
X-ray emission on all scales, which may be more easily appreciated 
in images that are less heavily smoothed (e.g., Fig. 8).

\begin{figure*}[htb!]
\centerline{ {\hfil\hfil
\hfil\hfil}}
\caption{\protect\footnotesize
ACIS-I diffuse X-ray intensity images in three bands: 0.5-1 keV 
(a), 1-2 keV (b), and 2-8 keV (c). The images
are adaptively smoothed with a Gaussian, the size of which is adjusted to 
achieve a uniform count-to-noise ratio of $\sim 6$ across the field.
Removed sources are marked with {\sl crosses}, whereas regions of the
large prominent diffuse features are labeled in (a)}
\label{fig8}
\end{figure*}

Let us first discuss the relatively small-scale, but quite conspicuous, 
diffuse X-ray feature, the southwest (Figs. 2, 7, and 8). 
Fig. 9 shows a close-up
of the field, which also contains three discrete sources
(Table 1; Fig. 2). X-17
is identified with a galaxy, which our optical spectroscopy shows to 
be at z=0.29 (Miller et al. 2004). The other two sources 
(X-9 and X-11) also appear to have optical
counterparts, albeit very faint. These discrete X-ray sources all have hard
spectral characteristics, as evidenced by their relative high fluxes in the
2-8 keV band (Fig. 7a). In comparison, the extended 
X-ray feature has a substantially softer spectrum (Fig. 10; \S 2), 
with the emission primarily in the 0.5-2 keV band. 

What might be the nature of the southwest? Located at a large
off-axis angle in the ACIS-I image, the feature could be
a combination of several physically-unrelated components. The 
X-ray counting statistics of the feature ($\sim 540$ counts) is also not
sufficient for a detailed
spectral analysis. Nevertheless, we have attempted 
a simple spectral characterization of the feature, by fixing the absorption
to be the Galactic foreground ($N_H \sim 3 \times 10^{20} {\rm~cm^{-2}}$).
A spectral fit with a thermal plasma model then suggests that the feature
may represent a cluster at $z \sim 1$ (Fig. 10).  The temperature 
3.1(2.3-4.1) and the inferred 0.5- 8 keV luminosity 
$\sim 1.4 \times 10^{44} {\rm~erg~s^{-1}}$ 
(in the rest frame of the putative cluster)
are consistent with the normal correlation between the two parameters 
expected for clusters 
(Wu, Xue, \& Fang 1999, corrected for our adopted Hubble constant). 
The small size of the feature also agrees
with this interpretation. But, the best-fit metal abundance
seems to be unusually high (greater than the solar; the upper limit is not
constrained), mainly due to two 
apparent emission line features at $\sim 1.0$ (Si{\small XIV}) and 1.2 (S{\small XV}-S{\small XIV}).
Furthermore, the X-ray feature is
apparently associated with an excess of very faint galaxies (Fig. 9), again
consistent with the distant cluster interpretation. Clearly, more 
observations are required to establish the true nature of the X-ray feature.
We will not discuss it further in the present work.

\begin{figure} 
\centerline{ {\hfil\hfil
\hfil\hfil}}
\caption{\protect\footnotesize
Overlay of ACIS-I intensity contours on a Kitt Peak 4-m R-band 
image of the southwest (see also Fig. 2).
The X-ray intensity is smoothed in the same way as in Fig. 7.
The contour levels are at 3.6, 3.9, 4.5, 5.4, 6.6, 8.1, 10, 18,
33, and 63  $\times 10^{-3} {\rm~counts~s^{-1}~arcmin^{-2}}$.
}
\label{fig9}
\end{figure}

\begin{figure} 
\centerline{ {\hfil\hfil
\psfig{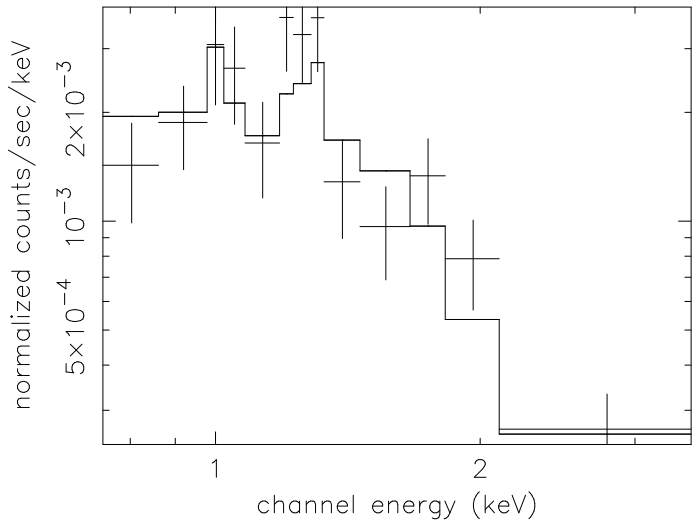}
\hfil\hfil}}
\caption{\protect\footnotesize
ACIS-I Spectra of the southwest, 
together with the best-fit thermal plasma spectra model
(see text for details. The on-source 
spectral data were extracted from the
ellipse outlined in Fig. 2 (see also \S 2).}
\label{fig10}
\end{figure}

The remaining diffuse X-ray emission is correlated 
well with identified complex member galaxies (e.g., Fig. 7b).
Two relatively prominent concentrations are sampled by the ACIS-I 
observation: the Abell 2125 cluster and the southwest LSBXE patch. The X-ray spectra 
of these two diffuse X-ray features are presented in Fig. 11. The total 
number of counts included in the spectral fits (Fig. 11) 
is 5627 for the cluster and 1819 for the LSBXE, including
about 1450 and 790 background counts within the respective circular regions defined in Fig. 2. The spectra are grouped to
achieve a net (background-subtracted) signal-to-noise ratio greater than 3
in each bin. The results of the spectral fits are summarized in Table 2. We
have obtained these results by assuming the mean optical 
redshift of the complex to be 
$z = 0.247$, which is consistent with our direct fit to
the X-ray spectral data: $z = 0.23(0.21-0.25)$ for \xs\ and $0.30(0.06-0.5)$
for the LSBXE (all uncertainty ranges are at the 90\% confidence level).  
The results are not sensitive to the exact 
aperture used for the spectral extraction (Fig. 2).

\begin{figure*}[tbh!]
\unitlength1.0cm
\centerline{
    \begin{picture}(17,8) 
\put(0,0){ \begin{picture}(8,8)
        \end{picture}
        }
\put(8.5,0){ \begin{picture}(8,8)
        \end{picture}
        }
    \end{picture}
}
\caption{\protect\footnotesize
ACIS-I Spectra of the \xs\ cluster (left panel) and the LSBXE 
(right panel), together with the best-fit spectra models (Table 2).
}
\label{fig11}
\end{figure*}

\begin{figure} 
\centerline{ {\hfil\hfil
\hfil\hfil}}
\caption{\protect\footnotesize
Characterization of the 2-D 
morphology of the 0.5-8 keV diffuse emission from the Abell 2125 
cluster with ellipses on evenly-spaced scales from 
0\farcm5 - 2\farcm5. The gray-scale map is in
the 0.5-2 keV band and is adaptively smoothed with a count-to-noise 
ratio of 6.}
\label{fig12}
\end{figure}

\begin{figure*} 
\centerline{ {\hfil\hfil
\psfig{figure=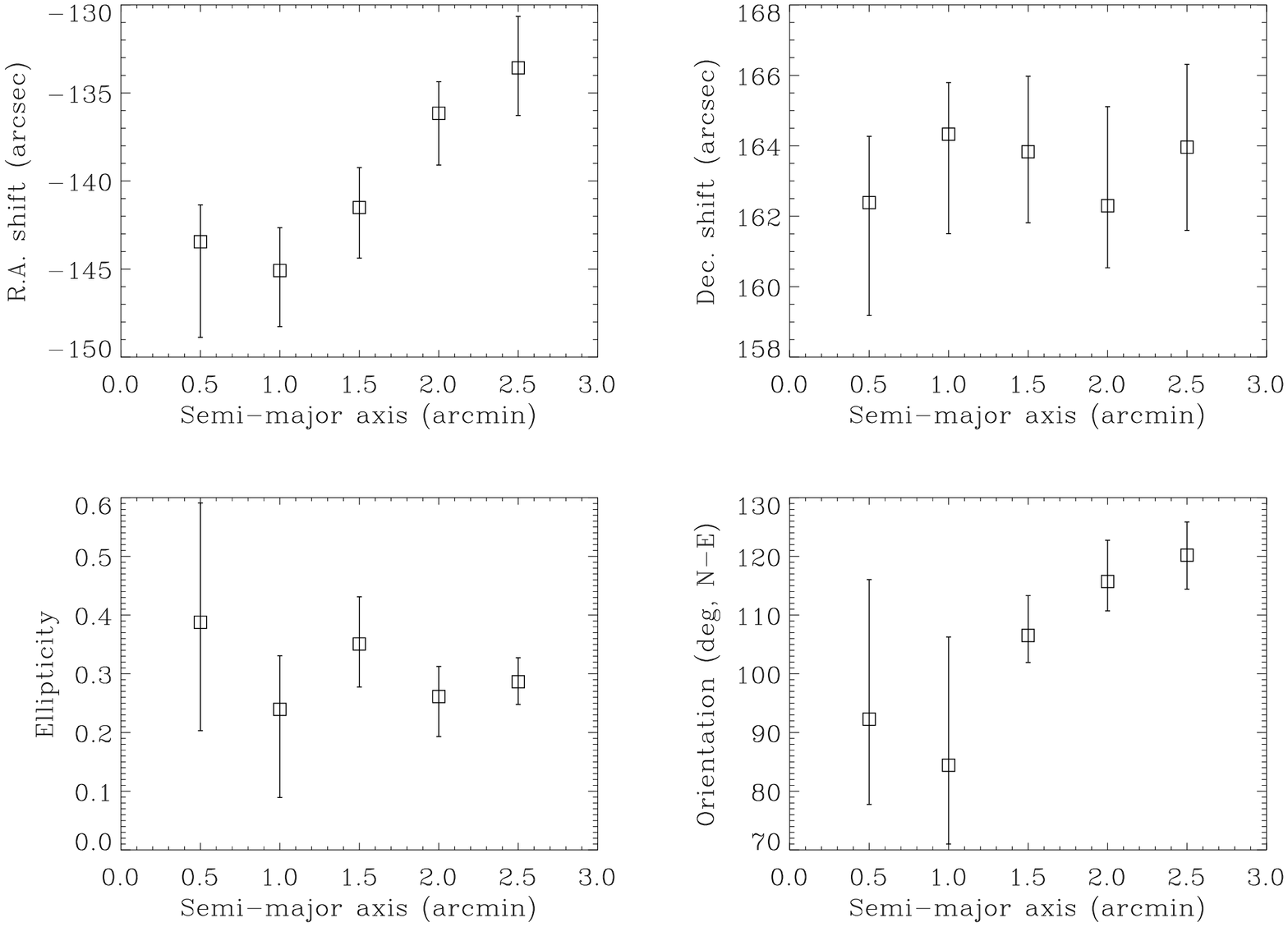,height=4.5in,angle=90, clip=}
\hfil\hfil}}
\caption{\protect\footnotesize
Center shifts, ellipticity, and orientation of the 0.5-8 keV 
intensity isophote
ellipse as a function of semi-major axis
of the Abell 2125 
cluster. Error bars are at the 90\% confidence level. 
 }
\label{fig13}
\end{figure*}

The diffuse X-ray surface brightness intensity of a relaxed 
cluster can typically be
characterized by the standard $\beta$-model of the form (Cavaliere \& Fusco-Femiano 1976):
\begin{equation} 
I = I_{o} \left(1+ {r^2 \over r_c^2}\right)^{1/2-3\beta},
\end{equation} 
where $r$ is the off-center radius. However, the model
gives a poor fit
to the azimuthally-averaged radial surface brightness intensity 
distribution in the 0.5-8 keV band ($\chi^2/d.o.f. = 107/56$) and
can be rejected at a confidence of $\sim 5\sigma$. 

The morphology of the cluster is far from the axis-symmetric.
We characterize the morphological variation of the cluster with a series 
of ellipses on various scales (Fig. 12). Each ellipse is determined by
the moments
of the enclosed X-ray intensity distribution (e.g., Carter \& Metcalfe 1980;
Wang, Ulmer, \& Lavery 1997b). We follow the iterative approach detailed in Wang et al. 
(1997b) to calculate the four parameters that define an ellipse: 
the center coordinates (R.A. and Dec. shifts relative to the aiming 
direction of the ACIS-I observation), ellipticity ($\epsilon$), and 
orientation of the major axis ($\theta$; north to east). The results 
obtained from the 0.5-8 keV band data are presented in Figs. 12 and 13. 
Note that
the 90\% statistical confidence error bars on different scales are not totally 
independent and thus overestimated, 
as the calculation of the moments uses all the data 
with each ellipse (Wang et a. 1997b).  The most significant change
of the ellipse parameters with 
the semi-major axis is the center R.A. position, which shifts
by $\sim 10^{\prime\prime}$ from the inner region to outer region of 
cluster. The centroid of the cluster is offset to the northwest 
from the central triple of
major cD-like galaxies (Fig. 7c). This offset as well as the centroid
shift and the elongation of the cluster morphology may be a manifestation of
the underlying gravitational mass distribution. Galaxies in the field 
show another concentration about 1\farcm4 to the northwest (Fig. 7c). But the diffuse X-ray emission extends further to the north on
larger scales (Fig. 12). This extension is reflected in the change of the 
ellipse orientation with the scale (Fig. 13, lower right panel).

\begin{figure} 
\centerline{ {\hfil\hfil
\psfig{figure=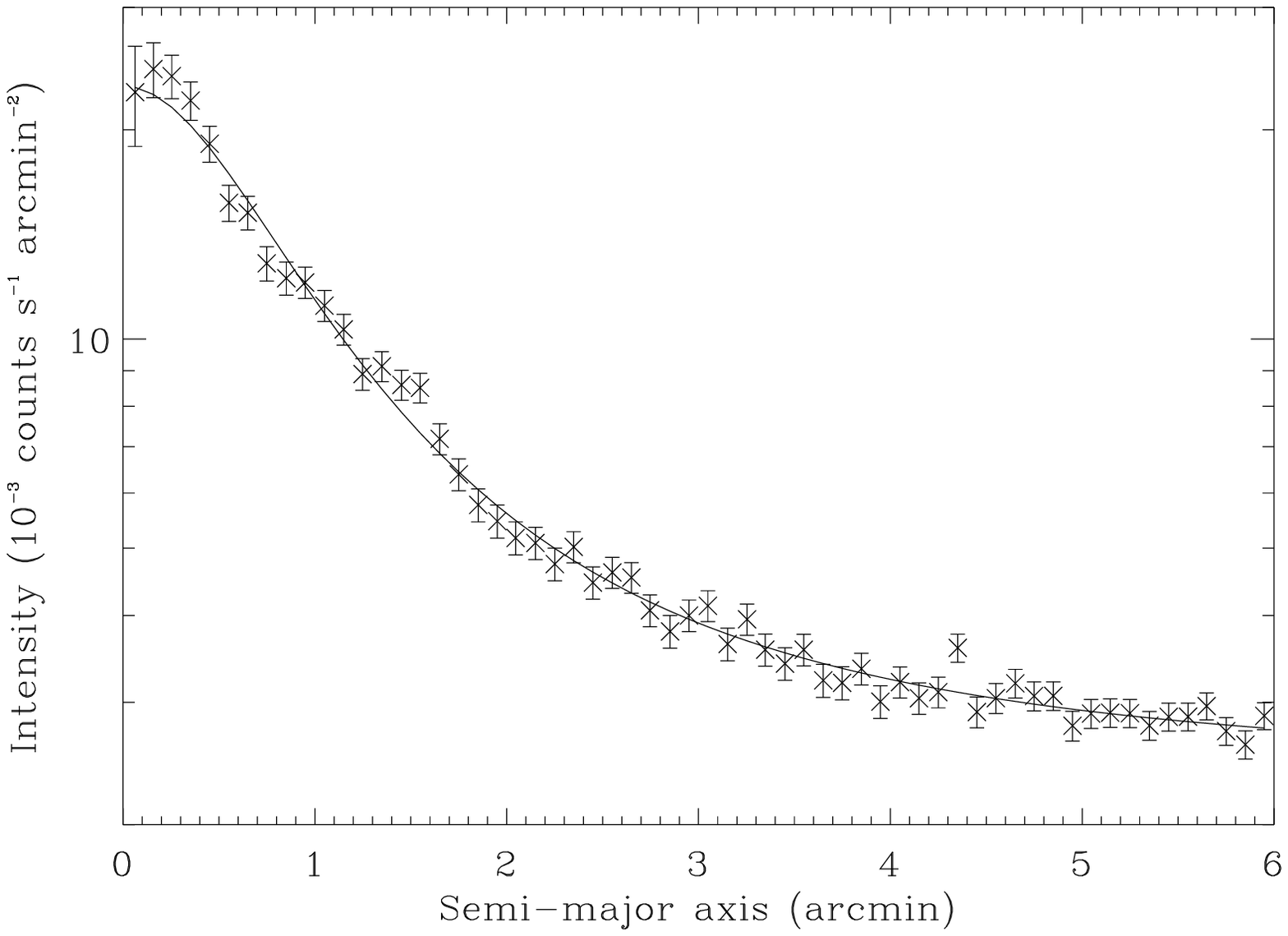,height=2.7in,angle=0, clip=}
\hfil\hfil}}
\caption{\protect\footnotesize
ACIS-I 0.5-8 keV surface brightness profile of Abell 2125 in the elliptical
coordinates defined by the best-fit isophote ellipse on the scale of 5$^\prime$
semi-major axis. The curve represents the best-fit $\beta$-model.}
\label{fig14}
\end{figure}

Ignoring these small, though significant, morphological variations, we use the 
elliptical coordinates as defined by the ellipse parameters on the 
2\farcm5 scale (Table 3) to measure
the surface brightness intensity profile as a function of the semi-major axis. 
A $\beta$-model fit 
to this profile obtained in the elliptical coordinates 
is acceptable (Fig. 14; Table 3) and 
may thus be considered
as a characterization of the average large-scale properties of the cluster.
We infer from the model the central electron density as $n_0 \sim 2.0 
\times 10^{-3} {\rm~cm^{-3}}$, 
assuming an oblate shape
of the X-ray-emitting medium and the best-fit thermal plasma model (Table 2).
The mean cooling time scale of the gas at the center of the cluster
is then $ \sim 18$ 
Gyr, which is longer than the age of the Universe.

We have also conducted the same morphological characterization 
of the cluster in the 0.5-2 keV band. The 
2-D ellipse fit in this band gives the ellipse parameters that are nearly
identical to those in the 0.5-8 keV band, especially on the 2\farcm5 scale. 
Except for the central intensity $I(0)$, the best-fit $\beta-$model 
parameters in the two bands are statistically consistent with each other, 
particularly with the consideration
that $\beta$ correlates with the core radius $r_c$ in the parameter
estimation. But 
the fit is much less satisfactory in the 0.5-2 keV band 
than in the 0.5-8 keV band.

\subsection{X-ray Substructure in the Central Cluster}

A careful inspection of the cluster images in various bands indicates that 
much of the deviation of the X-ray intensity distribution from the global
elliptical $\beta$-model is caused by the presence of small-scale 
substructures, which are particularly apparent in the 0.5-2 keV band. 
For example,
Figs. 7c,d clearly show energy-dependent features, which
are strikingly associated with several massive radio galaxies in the 
cluster. Comparison with point-like sources
in the vicinity indicates that these peaks are resolved. Their shapes 
in the figures are, however, affected by the overall X-ray intensity gradient 
across the cluster. To minimize this effect, we subtract the 
best-fit elliptical $\beta$-model from the 0.5-2 keV intensity map. 
Fig. 15 compares the X-ray intensity residuals  
and the radio contours. The enhanced soft X-ray peaks
are clearly associated with the galaxies 00047 (z=0.2528) and 00057 (z=0.2518),
although only the former is detected as a discrete source (Table 1).

\begin{figure} 
\unitlength1.0cm
\centerline{ 
}
\caption{\protect\footnotesize
Residual map of the ACIS-I intensity in the core of 
the Abell 2125 cluster, after subtracting the  best-fit 
elliptical $\beta$-model (Fig. 14; Table 3). The image is smoothed with a 
Gaussian of FWHM equal to 6$^{\prime\prime}$. 
The overlaid 20-cm continuum intensity contours are at 
0.04, 0.08, 0.16, 0.32, 0.64, 0.96, 2, 4, and 8 mJy/beam
(Dwarakanath \& Owen 1999).
The large thick {\sl plus} sign 
marks the adopted global cluster center (Table 3), while the {\sl crosses} 
show the locations of the optical positions of the four radio galaxies:
00047, 00057, 00105, and 00106 from right to left (Owen et al. 2004b). 
}
\label{fig15}
\end{figure}

\begin{figure} 
\centerline{ {\hfil\hfil
\psfig{figure=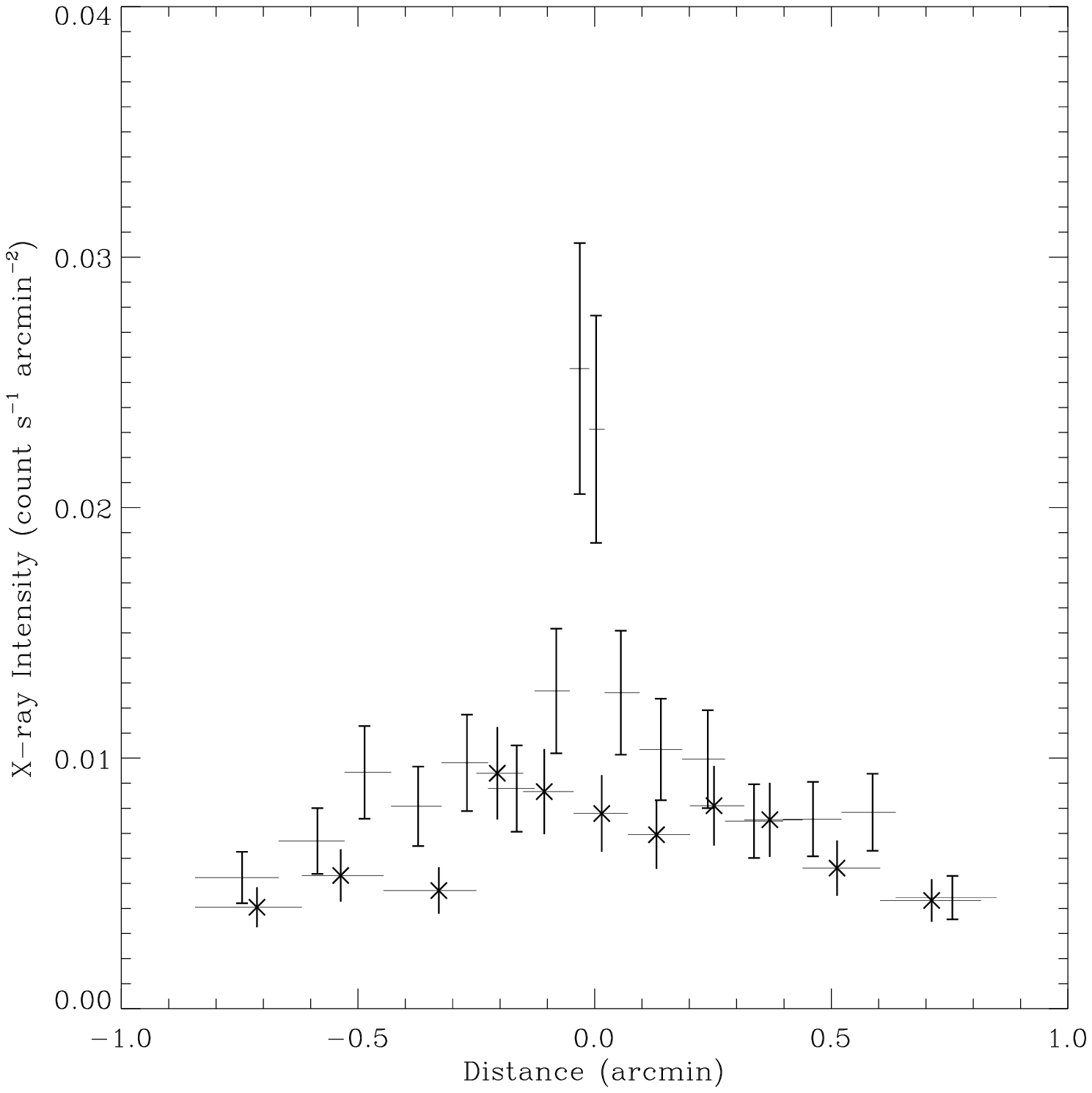,height=3.in,angle=0,angle=90, clip=}
\hfil\hfil}}
\caption{\protect\footnotesize
Average ACIS-I intensity distributions along a cut, which is 22'' wide
and perpendicular to the trail apparently following X-68 or the
optical/radio galaxy C153 (see also Figs. 7c,d and 15). The data 
points
are obtained in two bands: 0.5-1.5 keV and 1.5-4 keV (marked by ``x''). The
vertical error bars are at 1$\sigma$, whereas the horizontal bars represent the 
off-trail distance ranges of the data points.
}
\label{fig16}
\end{figure}

The most outstanding feature in this residual image is a ``trail'' 
(X-68) apparently attached to the optical galaxy 00047, 
which has a radio counterpart C153 (Owen et al. 2004a; Fig. 15; 
see also Figs. 7c,d). 
This trail is only apparent in the 0.5-2 keV band and extends about 
$22^{\prime\prime}$ (corresponding to $\sim 88$ kpc at the distance of 
the cluster) from the galaxy to the northeast. A similar, though shorter,
trail in the same direction is also seen in [OII] line emission 
(Owen et al. 2004b). With the
limited counting statistics of the data, however, we cannot conclude that
this apparent trail indeed represents a coherent diffuse X-ray structure.
While the association of the first $\sim 10^{\prime\prime}$ 
high surface brightness portion of the trail 
with the galaxy is convincing, the rest may become mixed up with
other things, including possible weak discrete sources and/or 
statistical fluctuations. 
Fig. 16 compares the surface brightness intensity across the trail.
From this comparison, we estimate the average width of the trail as 
$\sim 4^{\prime\prime}$ ($\sim 16$ kpc) and the total net count rate 
above the local background of $\sim 
0.01{\rm~ counts~s^{-1}~arcmin^{-2}}$ as $\sim 5.3(\pm 0.9) \times 10^{-4} 
{\rm~ counts~s^{-1}}$ in the 0.5-1.5 keV band. 

\section{Discussion}

The above results provide a detailed
characterization of the diffuse X-ray emission and discrete sources in
the Abell 2125 complex. In the following, we try to address several key 
issues about the complex by incorporating the results with information 
learned from observations in other wavelength bands. 

\subsection{Abell 2125 as a Large-scale Hierarchical Complex}

This distinct complex of galaxies and hot gas shows several
remarkable characteristics:

\begin{itemize}
\item a large velocity dispersion which appears to be
due to two or three major components  with a total projected extent of
several Mpc (Miller et al 2004);

\item an exceptionally large fraction of radio galaxies, which are located
primarily outside the central Abell 2125 cluster (Owen et al. 2004a);

\item the presence of multiple X-ray-emitting clusters and LSBXE 
features, each with substantially lower luminosity and temperature
than expected from the overall galaxy richness and 
large velocity dispersion of the complex (see also Wang et al. 1997).
\end{itemize}

Comparisons with various numerical simulations 
indicate that the complex represents a projection of multiple components, 
which might be in a process
of merging with each other (Miller et al. 2004; Owen et al. 2004a). 
Projection effects, together with enhanced
activities during this process, may explain the observed characteristics.

\subsection{Nature of the LSBXE}

Globally speaking, the prominent southwest patch of the 
LSBXE characterized here probably represents a local 
temperature and density enhancement of a large-scale diffuse 
intergalactic medium (IGM) structure, as 
indicated first in the \rosat\ PSPC image (Fig. 1), which is more sensitive to
very soft X-rays ($\lesssim 0.5$ keV) than the \chandra\ observation.

The LSBXE is distinctly different from 
the central cluster. Even the relatively X-ray-bright southwest patch
does not seem to be centrally peaked, as may be expected 
from a more-or-less virialized intracluster medium (ICM). The kinematics
(Miller et al. 2004) indicates that the galaxies 
in this patch are probably loosely bound, although 
their exact line-of-sight distribution is not clear.
Figs. 17c,d  show a sub-field of the LSBXE, as sampled by an {\sl HST} WFPC-2
image. X-14 and X-22 may be associated with two bright ellipticals in 
the field. But the centroid of X-14 is closer to a faint point-like optical
object about 4$^{\prime\prime}$ offset from the corresponding elliptical.
Therefore, the X-ray source may represent a background AGN.
Most of other galaxies in the field appear to be spirals. X-ray
emission from such galaxies are typically dominated by X-ray binaries and/or
AGNs and is therefore expected to have a relatively hard spectrum. 
But the diffuse X-ray enhancement 
associated with the galaxy concentration is soft, indicating 
an origin in diffuse hot gas around individual galaxies,
in groups of galaxies, and/or in the intergroup IGM.

The heavy element abundance in the LSBXE appears to be substantially lower 
than that in the ICM of the central cluster. The abundance may be 
underestimated in the LSBXE, if it contains multiple temperature components.
Although the single temperature thermal 
plasma gives a satisfactory fit, we also tried a model with 
two temperature components. This model gives a marginally improved
 fit to the LSBXE spectrum ($\chi^2/d.o.f = 18.2/23$, compared with 
that that in Table 2), but provides no
useful constraints on the
abundance, because of the limited counting statistics of the data.
If the low metallicity obtained from the single temperature plasma fit is 
real, there may be two possible explanations: 1) Metals (mainly irons) are 
largely locked in dust grains within or 
around galaxy groups (e.g., Nollenberg, Williams, \& Maddox, 2003); 
2) The metals are still bounded by individual 
galaxies (e.g., Tsai \& Mathews 1996). 
Dust grains may survive, and may even not mix well with, 
the X-ray-emitting gas in a 
quiescent, low density environment. But the mixing 
should likely to occur during or after the merger of a galaxy or 
group with a cluster. The dust grains would then be 
destroyed rapidly by sputtering in the ICM of high density and temperature.
This may explain the metal abundance difference in the
X-ray-emitting gas between the LSBXE and
the central cluster.

\subsection{Dynamical State of the Central  Cluster}

This central cluster (Abell 2125; Fig. 1) 
represents the strongest enhancement of the diffuse X-ray
emission in the complex. The relative values of our measured temperature 
and luminosity of the  ICM (Table 2) are consistent with those for typical 
clusters (e.g., Wu et al. 1999).
From the mass-temperature 
relationship (e.g., Shimizu et al. 2003), we may then expect the virial mass
of the cluster as $2-5 \times 10^{14} M_\odot$.  
For a virialized system, the temperature would also predict
a galaxy velocity dispersion of $\sim 6.5 \times 10^{2} 
{\rm~km~s^{-1}}$. This velocity dispersion is considerably smaller than 
$\sim 8.3 \times 10^{2} {\rm~km~s^{-1}}$ (assuming to be isotropic), 
inferred from the radial velocity dispersion of $\sim 4.8
\times 10^{2} {\rm~km~s^{-1}}$ estimated from the modeling of the 
velocity field (Miller et al. 2004). 
Interestingly, the two brightest compact radio galaxies,
00047 and 00057, are moving  
at high velocities ($\sim 1.5-1.8 \times 10^3 {\rm~km~s^{-1}}$) 
relative the mean of the cluster. It is conceivable that these two 
galaxies might simply represent a background cluster/group projected 
in the same sky by chance. The projected closeness of the galaxies 
to the center of the cluster and their enhanced diffuse X-ray emission,
however, suggest that the association is 
physical, consistent with that the cluster is experiencing a major 
merger probably in a direction close to the line of sight. Indeed, 
00047 is a strongly disturbed disk-like galaxy and the associated diffuse
X-ray emission forms a trail, as expected from a ram-pressure stripping.
00057 and other two radio galaxies in the core of the cluster are all cD-like
ellipticals (Figs. 7c and 15). 

Furthermore, the 2-D morphology of the cluster in the sky, as shown in \S 3.2,
also suggests that it
is far from a relaxed system. The X-ray morphology is strongly elongated and 
shows the centroid shift with scales, strongly indicating an ongoing merger
(e.g., Roettiger et al. 1993). The measured X-ray ellipticity of the
cluster is the largest in the Butcher \& Oemler sample (Wang \& Ulmer 1997). 
The X-ray centroid is further offset from the
cD galaxies in the cluster core (Fig. 15). A northwest-southeast elongation is also seen 
in the galaxy distribution, similar to the X-ray morphology (Owen et al
2004b). One separate sub-cluster, at least, can be found in the NW extension
of the cluster and is centered on a radio-loud cD-like galaxy (00039)
and coincides with the bright soft X-ray source X-65 (Fig. 17a). This galaxy has 
a similar 
optical magnitude as those three cDs in the core. Furthermore, the velocity of 
the galaxy
is comparable to those of the north and northeast cDs in the core, suggesting that the
merger axis is nearly perpendicular to the line of sight. 
This northwest-southeast sub-cluster merger provides a natural explanation for the
wide-angle tailed radio source attached to one (00106) of the two low-velocity 
ellipticals in the cluster core (Fig. 15), 
which suggests a pressure gradient or relative motion 
of the local gas in the plane of the sky. In short, the
central cluster is undergoing a merger, possibly in multiple directions.
As clusters are expected to be found at intersections of cosmic webs,
such a complicated merger process is probably common for  
clusters still in early formation stages. 

What might be the relationship between the LSBXE and the central cluster?
The velocity centroids of the associated galaxy concentrations 
appear to be comparable (Miller et al. 2004). Spatially, the two 
concentrations are separated by
 $\sim 7^\prime$, or a projected distance $d_p \sim $ 1.6 Mpc. One possibility 
is that the cluster and the LSBXE are merging in the plane of the sky.
There are galaxies occupying the area between the LSBXE and the 
central cluster with the similar velocity (as well as a separate higher
velocity component; Miller et al. 2004). There is also faint diffuse X-ray
emission in the area, although the quality of the data does not allow
for a distinction between a projection effect and a true dynamical 
interaction between these two concentrations. 
Alternatively, the two concentrations may be 
physically separated by a turn-around distance
$d_{t}$ so that their relative velocity is near zero. 
Following Sarazin (2003), we estimated $d_{t} \approx 
2(GM)^{1/3} (t_t/\pi)^{2/3}
\approx (7 {\rm~Mpc}) (M/10^{15} M_\odot)^{1/3} (t_t/10^{10} {\rm yr})^{2/3}$,
where $M$ is the total gravitational mass of the two concentrations,
and $t_t$ is the age of the Universe at the turn-around time. If 
the system starts to collapse for the first time at $z = 0.247$, the axis is
then oriented relative to the line of sight by an angle 
$\sim {\rm sin}^{-1}(d_p/d_t)
\sim 16^\circ$ for $M \sim 5 \times 10^{14} M_\odot$ (see above). 
If the two concentrations have already passed across each other 
(i.e., the first collapse occurred much earlier) and is about to 
re-collapse, the angle 
would then be larger. 

\subsection {Galaxy-Environment Interaction}

As marked in Table 1 and Fig. 7b, 10 X-ray sources
are identified as member galaxies of the Abell 2125 complex. Eight of them
are radio galaxies, preferentially in the LSBXE region. Most of these 
X-ray-loud galaxies are not resolved by {\sl Chandra}
and are probably dominated by AGN activities.
Two of the X-ray sources are apparently resolved and are
associated with the giant elliptical galaxy 00039 and the disturbed
disk-like galaxy 00047 (C153)
in the central Abell 2125 cluster (Figs. 15 and 17a,b). As our source detection algorithms
are optimized to detect point-like sources, extended sources
such as the one associated with 00057 
(Fig. 15) is not listed in Table 1. These extended soft X-ray enhancements
probably represent hot gas associated with individual massive galaxies or
groups of galaxies, which may have entered the cluster for the first time.
Outside the central cluster, the ambient density and relative velocity
are typically low and the ram-pressure stripping is
probably not important, at least for massive galaxies. The intergalactic
gas around galaxies may even cool fast enough to replenish 
the gas consumed for star formation (Bekki et al. 2002). 
As they are plugging into a cluster, the surrounding gas may then be 
compressed by the high ram-pressure of the ICM, resulting in
enhanced soft X-ray emission. The eventual stripping of the gas and dust 
may be important in both enriching the ICM and transforming the galaxies
(e.g., Bekki et al. 2002.

\begin{figure}[htb!]
\centerline{
}
\caption{\protect\footnotesize
{\sl HST} WFPC-2 V-band images of a field near the core of the Abell 2125 
cluster (a and b) and a field in the LSBXE (c and d). The
overlaid X-ray intensity contours are at 
2.50,  2.75,  3.25,  4.00,  5.00,  6.25,  7.75,  9.5,
      14.25, 26.75, and 51.75 for the 0.5-2 keV band (a); 
2.80,  3.20,  4.00,  5.20,  6.80,  8.80,  11.2,  14
      21.6,  41.6,  and 81.6  for the 2-8 keV band (b); 
1.55,  1.6,      1.7,      1.85,      2.05,      2.3,      2.65,      3.9,
      6.4,   and   11.4  for the 0.5-2 keV band (c); 
2.1,      2.2,      2.4,      2.7,      3.1,      3.6,      4.3,      6.8,
      11.8,   and   21.8  for the 2-8 keV band (d);
all in units of $10^{-3} {\rm~counts~s^{-1}~arcmin^{-2}}$. X-ray
source numbers (Table 1) are labeled. }
\label{fig17}
\end{figure}

C153 represents an extreme case of the ram-pressure stripping.
This galaxy probably went through the cluster central region 
quite recently (Fig. 15). The absence of the trail above 1.5 keV (Fig. 15) 
suggests that the gas 
in the trail is substantially cooler ($kT \lesssim 1.5$ keV) 
than the ambient ICM, assuming collisional ionization equilibrium. 
(The the gas, if 
heated from a cool phase, may well be out of the equilibrium.)
The emissivity of the hot gas in the ACIS-I 
0.5-1.5 keV band  (Fig. 15)
peaks at $kT \sim 0.7$ keV, but is within a factor of 2 as
long as $kT \gtrsim 0.35$ keV. We estimate the luminosity of the trail 
as $\sim 5 \times 10^{41} {\rm~ergs~s^{-1}}$ in the 0.5-2 keV band,
which only weakly depends on the assumed gas temperature.
To proceed further, we assume that the entire trail 
can be approximated as a uniform cylinder of $\sim 88$ 
kpc long and  16 kpc diameter (Figs. 15-16; \S 3.3). The mean electron density can then be estimated as $\sim 1.0 \times 10^{-2}
{\rm~cm^{-3}}\xi^{-0.5} $, where $\xi$ is 
the gas metallicity in the solar units. The dependence on $\xi$ assumes that
metal lines dominate the X-ray emission. If the X-ray-emitting gas in the 
trail is roughly in a pressure balance with the ICM 
($p/k \sim 7.7 \times 10^4 {\rm~cm^{-3}~K}$; \S 3.2),
the required gas temperature is then $kT \sim 0.6 {\rm~keV} \xi^{0.5}$.
We further estimate the total mass and  radiative
cooling time scale of the gas in the trail as $\sim 
5 \times 10^9 \xi^{-0.5} M_\odot$ and $ 
\sim 1$ Gyr. If the gas is far from uniform in the trail, however, the 
cooling time could then be much shorter and may be compared to the crossing 
time scale of the galaxy through the cluster (a few times $10^8$ years).

A more thorough multiwavelength investigation of galaxy properties 
and their relationship to the local and global environment will be 
presented by Owen et al. (2004a,b). In particular, our 
X-ray source detection reported here is optimized for point-like sources
and the detection threshold is quite conservative, which minimizes
the global probability of including spurious sources in our list. 
With the priori positions of individual galaxies, one can lower the 
threshold for a positive detection of their X-ray counterparts. Using a
threshold of $P = 10^{-3}$, for 
example, we have tentatively identified another 15 faint X-ray 
counterparts of the radio members of the Abell 2125 complex  (Owen et al.
2004a). These studies suggest that the Abell 2125 complex 
is in a special phase of 
the cluster formation, probably triggered by mergers among major 
components of a large-scale structure (e.g., Owen et al. 2004a).
The presence of an unusually large number of active galaxies
may be especially important for the heating of the IGM,
affecting the subsequent evolution of the complex. 

\section{Summary}

We have presented a deep \chandra\ ACIS-I observation of 
a large-scale hierarchical complex associated with the central \xs\
complex. The superb spatial resolution 
and broad energy coverage of this observation have allowed us for the first time
cleanly separate contributions from discrete sources and diffuse hot
gas in the complex. The main results and conclusions we have obtained
are summarized as follows:

\begin{itemize}
\item We have detected 99 discrete sources with an on-axis flux limit of 
$\sim 8 \times 10^{-16}$ ${\rm~erg~cm^{-2}~s^{-1}}$ in the 0.5-8 keV band. 
Ten of these sources are identified to have optical counterparts
spectroscopically confirmed as the complex members. X-ray sources outside
the central cluster appear to be point-like and have relative hard spectral
characteristics. Inside the cluster, X-ray sources/peaks are associated 
with giant radio galaxies and are both soft and extended, indicating an 
origin in diffuse hot gas. The remaining unidentified 
sources are statistically consistent
 with being interlopers.

\item We have characterized the global X-ray properties of the central
cluster. The ratio of the ICM temperature (3.2 keV) and 
the luminosity ($\sim 7.9 \times 10^{43} {\rm~ergs~s^{-1}}$ in 
the 0.5-2 keV band) of the central cluster appears normal. The central
cluster itself if not rich, consistent with its small 
X-ray-inferred size. The Abell radius encloses another 
cluster, plus parts of other larger scale structure of the complex. 
The ongoing mergers in the central cluster are apparent in its
morphology and kinematics. 
An X-ray morphology analysis
shows significant intensity isophote ellipse shift with the 
spatial scale. An elliptical $\beta$-model gives a satisfactory
fit to the surface brightness intensity profile in the 0.5-8 keV band, but
not in the 0.5-2 keV band, in which morphological distortion due to 
substructures becomes more important.

\item We confirm the detection of large amounts of low surface brightness
diffuse X-ray emission in regions beyond the central cluster. In particular, 
the gas in the relatively prominent southwest patch has a mean thermal 
temperature of $\sim 1$ keV and a heavy element abundance of $\lesssim 9\%$ 
solar, substantially smaller than that in the ICM of the cluster. The 
diffuse X-ray emission correlates with concentrations of galaxies,
most of which are spirals. This, together with the apparently low 
abundance, suggests that the gas represents the hot IGM associated with 
the large-scale hierarchical structure formation as predicted in theories.

\end{itemize}

\acknowledgements 
We thank David Smith for his help in exporting
the Log($N$)--Log($S$) analysis into XSPEC and the referee for
valuable comments. This work was funded by 
NASA under the grants GO1-2126 and NAG5--8999.

\vfil
\eject

\begin{deluxetable}{lrrrrrrrrrr}
\tabletypesize{\scriptsize}
  \tablecaption{{\sl Chandra} Source List \label{acis_source_list}}
  \tablewidth{0pt}
  \tablehead{
  \colhead{Source} &
  \colhead{CXOU Name} &
  \colhead{$\delta_x$ ($''$)} &
  \colhead{log(P)} &
  \colhead{CR $({\rm~cts~ks}^{-1})$} &
  \colhead{HR} &
  \colhead{HR1} &
  \colhead{HR2} &
  \colhead{Flag} \\
  \noalign{\smallskip}
  \colhead{(1)} &
  \colhead{(2)} &
  \colhead{(3)} &
  \colhead{(4)} &
  \colhead{(5)} &
  \colhead{(6)} &
  \colhead{(7)} &
  \colhead{(8)} &
 \colhead{(9)}
  }
  \startdata
   1 &  J153918.89+661700.8 &  2.1 &$ -12.4$&$     0.60  \pm   0.15$&                                       --& --& -- &  B, H    \\
   2 &  J153919.36+661629.6 &  2.1 &$ -13.1$&$     0.67  \pm   0.14$&                                       --& --& -- &  B, S    \\
   3 &  J153922.59+661825.4 &  1.0 &$ -20.0$&$     2.15  \pm   0.21$&             $-0.49\pm0.13$ & $ 0.51\pm0.10$ & -- &  B, S, H \\
   4 &  J153925.48+660848.2 &  3.1 &$  -7.5$&$     0.42  \pm   0.12$&                                       --& --& -- &  S       \\
   5 &  J153935.02+660525.8 &  1.8 &$ -20.0$&$     1.16  \pm   0.17$&                          --& $-0.09\pm0.18$ & -- &  B, S, H \\
   6 &  J153935.49+661455.9 &  0.7 &$ -20.0$&$     1.67  \pm   0.17$&  $-0.27\pm0.14$ & $ 0.50\pm0.12$ & $-0.69\pm0.15$ & B, S, H \\
   7 &  J153935.61+661244.8 &  1.7 &$ -11.3$&$     0.69  \pm   0.16$&                                       --& --& -- &  B, S    \\
   8 &  J153938.08+662102.7 &  1.7 &$ -15.7$&$     1.14  \pm   0.19$&                          --& $ 0.12\pm0.19$ & -- &  B, S, H \\
   9 &  J153945.15+660651.7 &  1.9 &$  -7.2$&$     0.78  \pm   0.15$&                                       --& --& -- &  S       \\
  10 &  J153945.28+661236.1 &  0.5 &$ -20.0$&$     2.53  \pm   0.28$&  $-0.38\pm0.14$ & $ 0.29\pm0.13$ & $-0.55\pm0.17$ & B, S, H \\
  11 &  J153946.05+660702.4 &  1.7 &$ -11.7$&$     0.62  \pm   0.13$&                                       --& --& -- &  B, H    \\
  12 &  J153946.67+662013.4 &  0.9 &$ -15.2$&$     2.32  \pm   0.21$&             $-0.67\pm0.10$ & $ 0.15\pm0.10$ & -- &  B, S, H \\
  13 &  J153949.30+661737.1 &  0.6 &$ -20.0$&$     2.40  \pm   0.21$&  $-0.58\pm0.10$ & $ 0.15\pm0.10$ & $-0.64\pm0.17$ & B, S, H \\
  14 &  J153952.87+660954.0 &  1.1 &$ -11.9$&$     0.43  \pm   0.10$&                                       --& --& -- &  r, o; B, S    \\
  15 &  J153957.97+661350.4 &  0.7 &$ -20.0$&$     0.84  \pm   0.13$&                          --& $-0.12\pm0.16$ & -- &  B, S    \\
  16 &  J153958.54+660808.5 &  1.6 &$ -10.6$&$     0.28  \pm   0.09$&                                       --& --& -- &  S, B    \\
  17 &  J154000.02+660551.2 &  0.3 &$ -16.1$&$    11.72  \pm   0.46$&  $-0.40\pm0.05$ & $ 0.43\pm0.04$ & $-0.34\pm0.07$ & B, S, H \\
  18 &  J154001.51+661518.2 &  1.0 &$ -10.0$&$     0.24  \pm   0.07$&                                       --& --& -- &  S, B    \\
  19 &  J154003.00+661527.1 &  1.1 &$ -20.0$&$     0.34  \pm   0.08$&                                       --& --& -- &  B, S    \\
  20 &  J154003.44+661532.8 &  0.6 &$ -16.7$&$     1.12  \pm   0.13$&             $-0.42\pm0.15$ & $ 0.44\pm0.14$ & -- &  B, S, H \\
  21 &  J154005.03+661016.5 &  1.2 &$  -7.1$&$     0.24  \pm   0.07$&                                       --& --& -- &  S       \\
  22 &  J154005.29+661012.5 &  0.9 &$ -13.2$&$     0.31  \pm   0.08$&                                       --& --& -- &   r, o; S, B   \\
  23 &  J154005.55+660718.5 &  1.3 &$ -16.8$&$     0.61  \pm   0.11$&                                       --& --& -- &  B, H, S \\
  24 &  J154008.12+661625.5 &  0.6 &$ -17.6$&$     0.71  \pm   0.11$&                          --& $ 0.25\pm0.18$ & -- &  B, S    \\
  25 &  J154009.13+661217.0 &  0.7 &$ -15.4$&$     0.32  \pm   0.08$&                                       --& --& -- &   r, o; S, B    \\
  26 &  J154009.20+661328.0 &  0.8 &$ -15.9$&$     0.28  \pm   0.07$&                                       --& --& -- &  B, H    \\
  27 &  J154010.07+661635.2 &  0.6 &$ -17.7$&$     1.06  \pm   0.13$&             $-0.52\pm0.15$ & $ 0.11\pm0.15$ & -- &  B, S, H \\
  28 &  J154010.75+661922.5 &  1.1 &$ -15.8$&$     0.55  \pm   0.10$&                                       --& --& -- &  B, S    \\
  29 &  J154012.39+661438.8 &  0.1 &$ -20.0$&$     8.57  \pm   0.37$&  $-0.52\pm0.05$ & $ 0.23\pm0.05$ & $-0.42\pm0.07$ & B, S, H \\
  30 &  J154012.55+660939.3 &  1.3 &$  -9.4$&$     0.23  \pm   0.07$&                                       --& --& -- &  B, H    \\
  31 &  J154012.77+661920.3 &  1.2 &$ -20.0$&$     0.49  \pm   0.10$&                                       --& --& -- &  B, H    \\
  32 &  J154014.83+661548.7 &  0.4 &$ -20.0$&$     1.38  \pm   0.15$&             $-0.52\pm0.13$ & $ 0.40\pm0.12$ & -- &  B, S, H \\
  33 &  J154016.54+661039.6 &  0.7 &$ -12.4$&$     0.27  \pm   0.07$&                                       --& --& -- &  r, o; B, S    \\
  34 &  J154017.83+661302.7 &  1.0 &$ -13.0$&$     0.27  \pm   0.08$&                                       --& --& -- &  B, S, H \\
  35 &  J154018.45+660712.6 &  0.7 &$ -20.0$&$     1.10  \pm   0.14$&                          --& $-0.03\pm0.15$ & -- &  B, S, H \\
  36 &  J154023.34+661651.1 &  1.1 &$ -12.4$&$     0.26  \pm   0.07$&                                       --& --& -- &  B, H    \\
  37 &  J154025.58+660838.4 &  0.3 &$ -16.4$&$     3.04  \pm   0.23$&  $-0.48\pm0.09$ & $ 0.04\pm0.09$ & $-0.34\pm0.13$ & B, S, H \\
  38 &  J154027.18+661050.6 &  0.4 &$ -16.9$&$     0.50  \pm   0.09$&                          --& $ 0.34\pm0.20$ & -- &  S, B    \\
  39 &  J154033.52+661908.4 &  1.0 &$ -20.0$&$     0.43  \pm   0.09$&                                       --& --& -- &  B, H    \\
  40 &  J154033.64+660800.7 &  0.7 &$ -20.0$&$     0.60  \pm   0.10$&                            --& --& $-0.20\pm0.18$ & B, H, S \\
  41 &  J154038.84+661125.8 &  0.7 &$ -10.9$&$     0.12  \pm   0.05$&                                       --& --& -- &  S, B    \\
  42 &  J154039.11+661006.8 &  0.5 &$ -20.0$&$     0.48  \pm   0.09$&                                       --& --& -- &  B, H, S \\
  43 &  J154039.46+661713.1 &  0.6 &$ -16.1$&$     0.78  \pm   0.11$&                          --& $ 0.61\pm0.14$ & -- &  B, S, H \\
  44 &  J154039.99+661236.6 &  0.2 &$ -16.5$&$     0.83  \pm   0.11$&                          --& $ 0.41\pm0.15$ & -- &  S, B, H \\
  45 &  J154043.73+660844.5 &  0.7 &$ -20.0$&$     0.62  \pm   0.11$&                                       --& --& -- &  B, S, H \\
  46 &  J154044.85+660904.1 &  0.8 &$ -16.1$&$     0.68  \pm   0.16$&                                       --& --& -- &  B, S    \\
  47 &  J154045.02+660507.1 &  1.2 &$ -20.0$&$     1.07  \pm   0.17$&                                       --& --& -- &  B, S, H \\
  48 &  J154045.34+661727.2 &  0.7 &$ -16.0$&$     0.38  \pm   0.08$&                                       --& --& -- &  B, S, H \\
  49 &  J154046.22+661053.3 &  0.8 &$ -10.2$&$     0.25  \pm   0.09$&                                       --& --& -- &  B, S    \\
  50 &  J154046.73+661320.9 &  0.3 &$ -20.0$&$     0.30  \pm   0.07$&                                       --& --& -- &  B, S, H \\
  51 &  J154048.03+662002.9 &  2.0 &$  -6.3$&$     0.27  \pm   0.08$&                                       --& --& -- &  B       \\
  52 &  J154048.55+661026.8 &  0.8 &$  -7.0$&$     0.11  \pm   0.05$&                                       --& --& -- &  B, H    \\
  53 &  J154048.86+661135.4 &  0.3 &$ -17.0$&$     0.44  \pm   0.09$&               $ 0.86\pm0.19$ & --& $-0.29\pm0.19$ & B, H    \\
  54 &  J154051.54+661414.2 &  0.5 &$  -6.5$&$     0.10  \pm   0.04$&                                       --& --& -- &  B       \\
  55 &  J154052.03+660632.4 &  1.7 &$  -7.0$&$     0.28  \pm   0.08$&                                       --& --& -- &  B, S    \\
  56 &  J154052.44+661236.9 &  0.1 &$ -20.0$&$     2.08  \pm   0.18$&  $-0.53\pm0.10$ & $ 0.17\pm0.10$ & $-0.47\pm0.15$ & B, S, H \\
  57 &  J154052.56+661424.9 &  0.5 &$  -8.2$&$     0.11  \pm   0.04$&                                       --& --& -- &  S       \\
  58 &  J154055.69+661458.6 &  1.3 &$  -6.4$&$     0.10  \pm   0.05$&                                       --& --& -- &  S       \\
  59 &  J154056.45+661628.6 &  0.1 &$ -20.0$&$     8.80  \pm   0.37$&  $-0.54\pm0.05$ & $ 0.33\pm0.05$ & $-0.26\pm0.08$ & B, S, H \\
  60 &  J154057.13+660917.8 &  0.7 &$ -12.9$&$     0.24  \pm   0.07$&                                       --& --& -- &  B, S    \\
  61 &  J154058.93+661742.6 &  0.9 &$ -11.1$&$     0.30  \pm   0.08$&                                       --& --& -- &  B, S    \\
  62 &  J154059.21+660640.0 &  1.3 &$ -11.6$&$     0.36  \pm   0.09$&                                       --& --& -- &  B       \\
  63 &  J154100.39+661903.0 &  1.5 &$ -10.4$&$     0.26  \pm   0.08$&                                       --& --& -- &  r, o; S, B   \\
  64 &  J154102.01+661721.4 &  0.3 &$ -20.0$&$     1.79  \pm   0.17$&  $-0.54\pm0.11$ & $ 0.20\pm0.11$ & $-0.01\pm0.19$ & o; B, S, H\\
  65 &  J154102.04+661626.5 &  0.3 &$ -20.0$&$     0.73  \pm   0.12$&                          --& $-0.25\pm0.16$ & -- &  r, o; B, S    \\
  66 &  J154102.76+661404.7 &  0.2 &$ -20.0$&$     0.77  \pm   0.11$&                          --& $ 0.48\pm0.16$ & -- &  S, B, H \\
  67 &  J154109.25+661448.7 &  0.5 &$ -12.9$&$     0.24  \pm   0.06$&                                       --& --& -- &  B, S    \\
  68 &  J154109.79+661544.7 &  0.5 &$  -8.9$&$     0.28  \pm   0.07$&                                       --& --& -- &  r, o; B, S    \\
  69 &  J154112.48+661717.0 &  0.9 &$ -15.5$&$     0.41  \pm   0.09$&                                       --& --& -- &  B, S    \\
  70 &  J154112.90+660502.8 &  2.0 &$  -9.6$&$     0.32  \pm   0.10$&                                       --& --& -- &  S, B    \\
  71 &  J154116.94+661627.2 &  0.8 &$  -8.7$&$     0.34  \pm   0.09$&                                       --& --& -- &  B       \\
  72 &  J154117.43+661923.8 &  1.4 &$ -16.2$&$     0.43  \pm   0.09$&                                       --& --& -- &  o; S, B    \\
  73 &  J154117.92+661343.0 &  0.4 &$ -20.0$&$     0.52  \pm   0.09$&                                       --& --& -- &  B, S, H \\
  74 &  J154120.83+660933.7 &  0.4 &$ -16.1$&$     1.71  \pm   0.18$&  $-0.23\pm0.13$ & $ 0.88\pm0.07$ & $-0.29\pm0.16$ & B, S, H \\
  75 &  J154121.93+661347.9 &  1.0 &$  -6.3$&$     0.12  \pm   0.05$&                                       --& --& -- &  H       \\
  76 &  J154123.79+662057.2 &  1.5 &$ -20.0$&$     0.60  \pm   0.12$&                                       --& --& -- &  B, H    \\
  77 &  J154127.33+661741.7 &  1.5 &$  -9.8$&$     0.39  \pm   0.10$&                                       --& --& -- &  B, S    \\
  78 &  J154127.43+661413.6 &  0.6 &$ -20.0$&$     0.63  \pm   0.11$&                          --& $ 0.79\pm0.15$ & -- &  B, S, H \\
  79 &  J154127.51+660637.6 &  0.8 &$ -20.0$&$     1.79  \pm   0.18$&  $-0.33\pm0.14$ & $ 0.38\pm0.12$ & $ 0.02\pm0.18$ & B, S, H \\
  80 &  J154128.35+661247.5 &  0.7 &$ -12.3$&$     0.25  \pm   0.07$&                                       --& --& -- &  B, S    \\
  81 &  J154132.46+660834.5 &  1.1 &$ -16.8$&$     0.73  \pm   0.14$&                          --& $ 0.75\pm0.15$ & -- &  B, S    \\
  82 &  J154133.15+661215.8 &  0.9 &$ -13.5$&$     0.30  \pm   0.08$&                                       --& --& -- &  S, B    \\
  83 &  J154133.78+661341.8 &  0.8 &$ -16.2$&$     0.40  \pm   0.08$&                                       --& --& -- &  B, S    \\
  84 &  J154133.86+660728.6 &  1.3 &$ -14.0$&$     0.44  \pm   0.10$&                                       --& --& -- &  B, S    \\
  85 &  J154137.33+661506.7 &  1.1 &$  -8.5$&$     0.28  \pm   0.08$&                                       --& --& -- &  B, S    \\
  86 &  J154141.29+660531.0 &  1.5 &$ -20.0$&$     0.95  \pm   0.15$&                          --& $ 0.35\pm0.17$ & -- &  B, S    \\
  87 &  J154143.47+661419.4 &  1.0 &$ -14.0$&$     0.31  \pm   0.08$&                                       --& --& -- &  r, o; B, S \\
  88 &  J154144.13+661848.7 &  1.3 &$ -13.9$&$     0.48  \pm   0.10$&                          --& $ 0.83\pm0.16$ & -- &  B, S    \\
  89 &  J154144.72+661143.0 &  1.6 &$  -8.6$&$     0.26  \pm   0.07$&                                       --& --& -- &  B, S    \\
  90 &  J154145.99+661038.3 &  1.6 &$  -8.9$&$     0.28  \pm   0.08$&                                       --& --& -- &  B, S    \\
  91 &  J154156.70+660738.9 &  1.9 &$ -10.7$&$     0.51  \pm   0.12$&                                       --& --& -- &  B, S    \\
  92 &  J154157.12+661211.6 &  0.5 &$ -17.5$&$     2.38  \pm   0.20$&  $-0.09\pm0.11$ & $ 0.86\pm0.07$ & $-0.31\pm0.12$ & B, S, H \\
  93 &  J154158.26+661312.5 &  1.1 &$ -15.9$&$     0.66  \pm   0.14$&                                       --& --& -- &  B, S    \\
  94 &  J154159.60+661514.9 &  1.4 &$ -12.2$&$     0.38  \pm   0.09$&                                       --& --& -- &  B, S    \\
  95 &  J154200.02+661843.8 &  2.0 &$ -10.5$&$     0.52  \pm   0.11$&                                       --& --& -- &  B, H    \\
  96 &  J154200.21+661053.0 &  1.8 &$  -7.2$&$     0.30  \pm   0.09$&                                       --& --& -- &  B, H    \\
  97 &  J154205.44+661604.4 &  1.0 &$ -20.0$&$     1.15  \pm   0.15$&               $ 0.68\pm0.13$ & --& $-0.46\pm0.13$ & B, H, S \\
  98 &  J154205.78+661529.1 &  1.9 &$  -9.6$&$     0.43  \pm   0.11$&                                       --& --& -- &  B, H    \\
  99 &  J154205.82+661629.8 &  1.6 &$ -16.5$&$     0.69  \pm   0.15$&                          --& $ 0.53\pm0.20$ & -- &  S, B    \\
\enddata
\tablecomments{ Column (1): Generic source number. (2): 
{\sl Chandra} X-ray Observatory (unregistered) source name, following the
{\sl Chandra} naming convention and the IAU Recommendation for Nomenclature
(e.g., http://cdsweb.u-strasbg.fr/iau-spec.html). (3): Position 
uncertainty (1$\sigma$) in units of arcsec. (4): The false detection probability P
that the 
detected number of counts may result from the Poisson fluctuation of the local 
background within the detection aperture [log(P) smaller than -20.0 is set 
to -20.0]. (5): On-axis (exposure-corrected) source count rate in the 
0.5-8 keV 
band. (6-8): The hardness ratios defined as 
${\rm HR}=({\rm H-S})/({\rm H+S})$, ${\rm HR1}=({\rm S2-S1})/{\rm S}$, 
and ${\rm HR2}=({\rm H2-H1})/{\rm H}$, 
where S1, S2, H1, and H2 are the net source count rates in the 
0.5--1, 1--2, 2--4, and 4--8~keV 
 bands, respectively, while S and H
represent the sums, S1+S2 and  H1+H2. The hardness ratios are calculated 
only for sources with individual signal-to-noise ratios greater than 4 
in the broad band (B=S+H), and only the values with 
uncertainties less than 0.2 are included.
(9): The label ``B'', ``S'', or ``H'' mark the band in 
which a source is detected; the detection with the most accurate position, 
as adopted in Column (2), is marked first. The label ``r'' and ``o'' denote
the identification as the Abell 2125 complex member galaxies 
in radio and optical. 
}
  \end{deluxetable}

\begin{deluxetable}{lll}
\tabletypesize{\footnotesize}
\tablecaption{Results of Spectral fits \label{spectrum}}
\tablewidth{0pt}
\tablehead{
\colhead{Parameter} & 
\colhead{Cluster} & 
\colhead{LSBXE}}
\startdata
\noalign{\smallskip}
Best-fit $\chi^2$/d.o.f.\dotfill  &                     155/151 & 22/25\\
\noalign{\medskip}
Column Density  ($10^{20} {\rm~cm}^{-2}$)\dotfill  &    1.6($\lesssim 3.7$) & 0.1 ($\lesssim 10$) \\
Abundance       ($10^{-2}$ solar)  \dotfill     &       0.24(0.12-0.39) & 0.03($\lesssim 0.09$)\\
Temperature     (keV)\dotfill                    &      3.2(2.8-3.7)&0.98(0.71-1.2)\\
Integrated EM   ($10^{10} {\rm~cm}^{-5}$) \dotfill &    5.8(5.3-6.4) &4.0(3.1-7.6)\\ 
$L_{\rm 0.5-8 keV} (10^{43} {\rm~erg~s^{-1}}$) & 7.9 & 2.0\\
\enddata
\tablecomments{The uncertainty ranges of the parameters are included in
the parentheses and are all at the 90\% confidence.
The integrated emission measure is defined as $1/4\pi \int_{\Omega} {\rm EM} d\Omega$,
where the integration is over a solid angle over which the data are collected and
${\rm EM}=\int n_e^2 dr$ with $n_e$ being the electron density (all units in cgs).}
\end{deluxetable}

\begin{deluxetable}{lcccc}
\tabletypesize{\footnotesize}
\tablecaption{$\beta$-model fits to the Cluster Surface Brightness Profiles}
\tablewidth{0pt}
\tablehead{
\colhead{Parameter} & 
\colhead{0.5-8 keV band} & 
\colhead{0.5-2 keV band}}
\startdata
\noalign{\smallskip}   
Center R.A. (J2000) &15$^h41^m$11\farcs3 &15$^h41^m$11\farcs6\\
Center Dec. (J2000) &66$^\circ 16^\prime 2^{\prime\prime}$ &66$^\circ 16^\prime 3^{\prime\prime}$\\
Ellipticity &0.29 & 0.29\\
Position angle ($^\circ$) &123 &120\\
$\beta$         & 0.52 (0.47-0.58) & 0.62 (0.56-0.73)\\
$r_c$ ($^\prime$)&0.910 (0.773-1.08)  &1.20(1.04-1.46) \\
\ \ \ (kpc) & 210(178 - 248)&276 (240-336)\\
$I_{o} (10^{-2} {\rm~cts~s^{-1}~arcmin^{-2}}$) &2.1(1.9-2.2)&1.5(1.3-1.6)\\
$\chi^2/d.o.f$  & 66/56 & 96/56\\
\enddata
\tablecomments{Uncertainties in parameter values,
as presented in parentheses, are all at the 90\% confidence level. 
}
\end{deluxetable}

\end{document}